\documentclass[runningheads]{llncs}
\usepackage{graphicx}
\usepackage{xcolor}
\usepackage[multi-part-units=single]{siunitx}
\usepackage{amsmath}
\usepackage{amssymb}
\usepackage{bm}
\usepackage{url}
\usepackage{arydshln}
\usepackage{booktabs}

\begin{document}
\title{Leveraging the Mahalanobis Distance to enhance Unsupervised Brain MRI Anomaly Detection}

\titlerunning{Mahalanobis Distance for Unsupervised Brain MRI Anomaly Detection}
%
\author{Finn Behrendt\inst{1} \orcidID{0000-0001-7191-6508} \and Debayan Bhattacharya \inst{1} \and Robin Mieling \inst{1} \orcidID{0000-0003-0262-2519} \and Lennart Maack \inst{1} \and Julia Krüger \inst{2} \and Roland Opfer \inst{2} \orcidID{0000-0002-9911-5478} \and Alexander Schlaefer \inst{1} \orcidID{0000-0001-9201-8854} } 
\authorrunning{F. Behrendt et al.}
%
\institute{ Institute of Medical Technology and Intelligent Systems, Hamburg University of Technology, Hamburg, Germany \and
Jung Diagnostics, Hamburg, Germany
 }
%
\maketitle              
\begin{abstract}
Unsupervised Anomaly Detection (UAD) methods rely on healthy data distributions to identify anomalies as outliers. In brain MRI, a common approach is reconstruction-based UAD, where generative models reconstruct healthy brain MRIs, and anomalies are detected as deviations between input and reconstruction. However, this method is sensitive to imperfect reconstructions, leading to false positives that impede the segmentation. To address this limitation, we construct multiple reconstructions with probabilistic diffusion models. We then analyze the resulting distribution of these reconstructions using the Mahalanobis distance to identify anomalies as outliers. By leveraging information about normal variations and covariance of individual pixels within this distribution, we effectively refine anomaly scoring, leading to improved segmentation. 
Our experimental results demonstrate substantial performance improvements across various data sets. Specifically, compared to relying solely on single reconstructions, our approach achieves relative improvements of 15.9\%, 35.4\%, 48.0\%, and 4.7\% in terms of AUPRC for the BRATS21, ATLAS, MSLUB and WMH data sets, respectively.

\end{abstract}
\section{Introduction}
Deep learning (DL) methods show promise in tasks like the segmentation of brain pathologies in magnetic resonance imaging (MRI) scans \cite{Lundervold.2019}. However, supervised DL methods require pixel-level annotations for training. This requirement becomes a challenge, particularly for screening tasks, where any pathology has to be detected even if not represented in the training data.
 Unsupervised Anomaly Detection (UAD) offers an alternative approach by learning the distribution of healthy data and identifying anomalies as outliers. A prevalent strategy is using reconstruction-based techniques \cite{Baur.2021b}. These methods train generative models (GM) on a data set composed solely of healthy brain MRI scans. The underlying assumption is that the GMs will fail to reconstruct anomalies or pathological structures not present in the training data set. Therefore, anomaly maps for segmenting abnormal structures can be derived from the deviations between input and reconstruction.
 However,  a critical challenge UAD methods face lies in their high sensitivity to errors stemming from imperfect reconstructions \cite{Meissen.2022c,Mao.2020,Cai.2023}. As a result, even healthy structures exhibit deviations in the anomaly map. Therefore, discriminating deviations caused by genuine pathologies from those arising due to imperfect reconstructions becomes challenging, leading to false positives in the final segmentation. 
While deviations from imperfect reconstructions are inevitable, analyzing multiple reconstructions of the same input can offer valuable insights into the normal variations within the distribution of pseudo-healthy reconstructions, potentially simplifying the discrimination. These multiple reconstructions can be sampled using probabilistic GMs. Previous approaches have primarily focused on comparing the average reconstruction with the corresponding input image \cite{Baur.2020,Baur.2021b}. However, these approaches ignore the valuable information in the variance and covariance of pixels across different reconstructions.
The inter-pixel covariance across reconstructions quantifies the relationship between pixel values at different locations. It can be utilized to achieve a more balanced decision when measuring the distance of individual input pixels to the pseudo-healthy distribution of healthy pixels.
Therefore, we propose using the Mahalanobis distance (MHD) \cite{Mahalanobis.1936} to measure the divergence of pixels in the input image from the pseudo-healthy distribution of pixels across multiple reconstructions. We employ denoising diffusion probabilistic models (DDPM) \cite{Ho.2020} to generate a pseudo-healthy reference distribution of reconstructions based on an individual input image. We then calculate the MHD between the input and the pseudo-healthy distribution to refine anomaly scoring. By considering the MHD in the pixel space with a full covariance matrix, we account for inter-pixel covariance. This enables capturing spatial information of neighboring pixels and long-range dependencies across pixels, such as symmetries in the reconstructions.
Our results indicate that refining anomaly scoring by the MHD can substantially enhance the segmentation performance of conditioned DDPMs (cDDPMs), particularly when considering the inter-pixel covariance of the generated pseudo-healthy distributions.
Compared to cDDPMs relying on single reconstructions, our approach leads to relative improvements of 15.9\%, 35.4\%, 48.0\%, and 4.7\% considering the AUPRC for the BRATS21, ATLAS, MSLUB and WMH data sets, respectively.
\subsection{Recent Work}
For most reconstruction-based approaches, AEs and VAEs are employed as GMs. While these architectures are conceptually simple and show promise in capturing the underlying distribution of healthy training data, their reconstructions tend to be blurry \cite{Baur.2021b}, substantially mitigating the segmentation performance. Therefore, many approaches aim to improve the reconstruction quality by focusing on spatial context \cite{Zimmerer.2019b} or utilizing 3D information \cite{Bengs.2021}. Also, vector quantized VAEs and soft intro VAEs are applied to UAD in brain MRI \cite{Pinaya.2022,Bercea.2023MIDL,Bercea.2023MICa}. 
Recent studies have indicated the effectiveness of DDPMs for UAD in brain MRI \cite{Pinaya.2022b,Wyatt.2022,Behrendt.2023,Behrendt.2023c}. Overall, GMs applied to the UAD task have shown promising progress. However, a crucial requirement for reconstruction-based UAD methods is to reconstruct healthy anatomy while avoiding the trivial replication of the input image. This necessitates the regularization of GMs, such as through a bottleneck in the latent space or additional regularization tasks like dropout \cite{Baur.2020} or denoising \cite{Kascenas.2022b}. Consequently, imperfect reconstructions become inevitable. 
However, probabilistic GMs offer the appealing property of sampling multiple reconstructions. The assessment of multiple reconstructions could add valuable information for discriminating anomalies from imperfect reconstructions in the anomaly map. However, only a few studies have explored using VAEs or Bayesian AEs with Monte Carlo dropout to sample multiple reconstructions \cite{Baur.2020,Baur.2021b}. These studies primarily concentrate on the mean of the generated reconstructions, which has not been shown to improve performance. Other approaches utilize uncertainty estimation to normalize the anomaly map by the estimated variance of individual pixels \cite{Sato.2019,Mao.2020,Cai.2023}. While this approach can improve the segmentation performance, it does not explicitly consider covariance across pixels. However, inter-pixel dependencies could provide valuable insights for anomaly scoring. Therefore, in this work, we focus on the inter-pixel dependencies and variations across different pseudo-healthy reconstructions and employ the MHD to measure the deviation of input pixels from the distribution of pixels in healthy reconstructions.
While the MHD is commonly used for outlier detection, its typical application is at the sample level within some aggregated feature space for sample-level anomaly detection \cite{Lee.2018,Vasiliuk.2023}. Furthermore, Saase et al. \cite{Saase.2020} apply the MHD in the pixel space using a healthy data set as a reference distribution, suggesting that simple statistical methods can compete with deep learning models. However, individual brains in the training data exhibit substantial differences. As a result, relying solely on these general population-based distributions could lead to a mismatch between individual test cases and the reference distribution, potentially impeding the segmentation.
\section{Methods}
We propose utilizing DDPMs to construct pseudo-healthy distributions specific to each individual case during evaluation. Subsequently, these case-specific distributions are employed as a reference to compute the Mahalanobis distance (MHD) in the pixel space to refine anomaly scoring.
\subsection{Generating Pseudo-healthy Distributions with DDPMs}
DDPMs are specialized in learning the distribution of training images $\bm{x} \in \mathbb{R}^{H,W,C}$, where $H$, $W$, and $C$ represent the height, width, and the number of channels, respectively. The training involves two primary processes: a forward process and a backward process.
In the forward process, an image $\bm{x}_0$ is incrementally transformed into Gaussian noise, represented as $\bm{x}_T = \bm{\epsilon} \sim \mathcal{N}(\textbf{0},\textbf{I})$. This transformation is guided by a predetermined noise schedule [$\beta_1,...,\beta_T$]. The intermediate image states $\bm{x}_t$ are generated by
\begin{equation*}
\bm{x}_t \sim q(\bm{x}_t|\bm{x}_0)=\mathcal{N}(\sqrt{\bar\alpha_t} \bm{x}_0,(1-\bar\alpha_t) \textbf{I}) \hspace{5mm} \bar\alpha_t=\prod\nolimits_{s=0}^{t}(1-\beta_t).
\end{equation*}
The noise level at each time step $t \in [1,...,T]$, influences $\bm{x}_t$, which can vary from being the original image (at $t=0$) to complete noise (at $t=T$).
In the backward process, the reconstruction of the original image $\bm{x}_0^{rec}$ from the noisy state $\bm{x}_T$ is given by
\begin{equation*}
\bm{x}_{0}^{rec} \sim p(\bm{x}_T)\prod\nolimits^T_{t=1}p_{\theta}(\bm{x}_{t-1}|\bm{x}_t) \hspace{5mm} p_{\theta}(\bm{x}_{t-1}|\bm{x}_t)=\mathcal{N}(\bm{\mu}_{\theta}(\bm{x}_t,t),\bm{\Sigma}_{\theta}(\bm{x}_t,t)).
\end{equation*}
Following \cite{Ho.2020}, $\bm{\mu}_{\theta}$ is estimated using a Unet \cite{Ronneberger.2015}, and $\bm{\Sigma}(t)$ is set to $\frac{1-\alpha_{t-1}}{1-\alpha_t} \beta_t \textbf{I}$. The training process entails minimizing the variational lower bound, which is approximated by the straightforward objective of predicting the added noise $\bm{\epsilon}{\theta}(\bm{x}t,t)$, as demonstrated in \cite{Ho.2020}. This yields the simplified loss function 
\begin{equation*}
\mathcal{L}_{simple} = ||\bm{\epsilon} - \bm{\epsilon}_{\theta}(\bm{x}_t,t)||^2.
\end{equation*}
In the context of reconstruction-based UAD, our objective is not to create new images from pure noise but to reconstruct healthy brain anatomy given an input image. Therefore, during testing, $\bm{x}^{rec}_{0}$ is estimated from $\bm{x}_t$, determining the extent of noise in $\bm{x}_t$ by $t_{test}<T$. 
To generate a distribution of multiple reconstructions, we sample $N$ versions of $\bm{x}_t$ by repeatedly resampling the additional noise and reconstructing each noised image by the denoising network. As we train the model on healthy data, this leads to a pseudo-healthy distribution consisting of $N$ different reconstructions of the given input image.
\subsection{Anomaly Scoring using Pseudo-Healthy Distributions and Mahalanobis Distance}
Our goal is to leverage the informative variations within the pseudo-healthy distribution of reconstructions. \\\\
\textbf{Averaged Reconstructions}
Initially, we calculate the mean reconstruction from multiple pseudo-healthy samples, represented as:
$\bm{\mu} = \frac{1}{N} \sum_{i=1}^{N} \bm{x}_i^{rec}$
Here, \(\bm{x}_i^{rec}\) denotes the, \(i\)-th pseudo-healthy reconstruction, and \(N\) represents the total number of reconstructions. The anomaly score is defined as the inverted pixel-wise Structural similarity index measure (SSIM) between the input image \(\bm{x}\) and the mean reconstruction \(\bm{\mu}\):
\begin{equation}
S_{mean}(\bm{x},\bm{\mu}) = 1 - SSIM(\bm{x},\bm{\mu}).
\end{equation}
Note that we use the pixel-wise SSIM as it has been shown to improve the anomaly scoring compared to intensity-based metrics \cite{meissen2022FAE,Lagogiannis.2023}\\\\
\textbf{Mahalanobis Distance}
The MHD is a statistical measure, quantifying the distance of a sample point from a multivariate reference distribution, considering its covariance. 
Employing the pixels of an input image as sample points that are compared to the pseudo-healthy distribution of reconstructed pixels, we can capture the degree of deviation of each pixel in the input image from what is 'expected' in the distribution of pseudo-healthy reconstructions. 
First, we start by calculating the MHD with a diagonal covariance matrix $\bm{\Sigma}_{diag} = diag(\bm{\sigma}^2) \in \mathbb{R}^{H\cdot W \times H\cdot W}$, where $\bm{\sigma}^2$ is the variance of each pixel across the $N$ reconstructions: $\bm{\sigma^2} = \frac{1}{N-1} \sum_{i=1}^{N} (\bm{x}_i^{rec}-\bm{\mu})^2$. Note that $\bm{x}$ and $\bm{\mu}$ are flattened to a dimension of $\mathbb{R}^{H\cdot W}$. This yields
\begin{equation}
MHD_{diag}(\bm{x}) = \sqrt{(\bm{x} - \bm{\mu})^\top \bm{\Sigma}_{diag}^{-1} (\bm{x} - \bm{\mu})}.
\end{equation}
This approach represents a standardization and allows for scaling the distance between input pixels and the mean reconstruction by the variance of individual pixels across different reconstructions. However, the diagonal covariance matrix does not consider covariance across different pixels.
To capture inter-pixel correlations, we extend our analysis to utilize a full covariance matrix, calculated as \\
$\bm{\Sigma}_{full} = \frac{1}{N-1} \sum_{i=1}^{N} (\bm{x}_i^{rec} - \bm{\mu})(\bm{x}_i^{rec} - \bm{\mu})^\top$ with dimension $\mathbb{R}^{H\cdot W \times H\cdot W}$,
leading to
\begin{equation}
MHD_{full}(\bm{x}) = \sqrt{(\bm{x} - \bm{\mu})^\top \bm{\Sigma}_{full}^{-1} (\bm{x} - \bm{\mu})}.
\end{equation}
After reshaping the MHD map to the input image shape, the final anomaly map is obtained by a per-pixel multiplication of the MHD map with the initial anomaly map for $S_{MHD} = S_{mean} \cdot MHD_{diag}$, and $S_{sMHD} = S_{mean} \cdot MHD_{full}$, respectively.
\subsection{Data}
Following the principle of reconstruction-based UAD, we utilize data sets without pathologies for training while evaluating data sets that contain annotated pathologies. \\
For training, we utilize the IXI data set \cite{BiomedicalImageAnalysisGroup.}, consisting of MRI scans in both T1- and T2-weighting. We split the training set into a healthy test set (N=160) and partition the remaining samples into 5 training sets (N=358) and 5 validation sets (N=44) for cross-validation. \\For evaluation, we utilize four different data sets, namely the BRATS21 \cite{Baid.2021} (N=1152), MSLUB \cite{lesjak2018novel} (N=30), ATLAS \cite{Liew.2022} (N=655) and WMH \cite{kuijf2019standardized} (N=60) data sets that exhibit tumors, multiple sclerosis, Stroke and white-matter lesions as pathologies, respectively. Note that while we train on both weightings separately, we evaluate BRATS and MSLUB on T2-weightings and ATLAS and WMH on T1-weightings.
Pre-processing of the data includes resampling to a voxel dimension of $1 \times 1 \times 1$ mm, skull-stripping, registration to the SRI ATLAS and N4 bias-correction. Furthermore, we crop 15 top and bottom slices and reduce the dimension by a factor of 2, leading to a resolution of $192 \times 192 \times 50 $ voxels. During training, we process the volumes slice-wise, with slices sampled with replacement. During evaluation, we iteratively reconstruct all slices to obtain the full volume.
\subsection{Implementation Details}
In this work, we build upon cDDPMs proposed in \cite{Behrendt.2023c} as a probabilistic GM. Compared to DDPMs, cDDPMs utilize an additional feature representation of the input image to guide the denoising process. We follow the architectural design of \cite{Behrendt.2023c} with a 3-layer Unet with channel dimensions [128, 128, 256] as a denoising network.  We calculate the SSIM anomaly score with a Gaussian kernel with a standard deviation of 1.  When calculating the MHD, we add a small regularization term (1e-5) to the diagonal entries of $\Sigma_{full}$ To ensure numerical stability during inversion. Additionally, we apply a Gaussian filter to the MHD map with a standard deviation of 1.
We compare established state-of-the-art baselines for UAD in brain MRI, including AE \cite{Baur.2021b}, VAE \cite{Baur.2021b}, DAE \cite{Kascenas.2022b}, DDPM \cite{Wyatt.2022}, pDDPM \cite{Behrendt.2023} and cDDPM \cite{Behrendt.2023c} as reconstruction-based approaches. Moreover, we compare RD \cite{Deng.2022} and FAE \cite{meissen2022FAE} as feature-based methods and the self-supervised approaches PII \cite{Tan.2021} and DRAEM \cite{Zavrtanik.2021}. Finally, we evaluate the covariance model (CM) of \cite{Saase.2020}, where the MHD is calculated with the healthy training set as a reference distribution. For AEs and VAEs, we set the latent dimension to 128. For VAEs, $\beta_{KLD} = 0.001$ is chosen. We train for 1600 epochs, using the ADAM optimizer, a learning rate of 1e-4 and a batch size of 32. For all DDPM-based models, we utilize simplex noise as introduced on \cite{Wyatt.2022}. We uniformly sample noise levels $t \in [1,T]$ with $T=1000$ during training and set the noise to  $t_{test} = \frac{T}{2} = 500$ during evaluation. All models are implemented in PyTorch v1.10 and trained on an NVIDIA A6000 graphics card\footnote{Code available at \\ \url{github.com/FinnBehrendt/Mahalanobis-Unsupervised-Anomaly-Detection}}.
For evaluation, we utilize the best possible Dice-Coefficient ($\lceil$Dice$\rceil$) and the Area under the precision-recall curve (AUPRC). 
Additionally, we employ the permutation test from the MLXtend library \cite{raschkas_2018_mlxtend}. This test involves 10,000 rounds of permutations and a significance level set at $\alpha = 5\% $ to assess statistical differences.
\section{Results}
\begin{figure}[t]
    \centering
    \begin{tabular}{cc}
    \includegraphics[width=.66\linewidth]{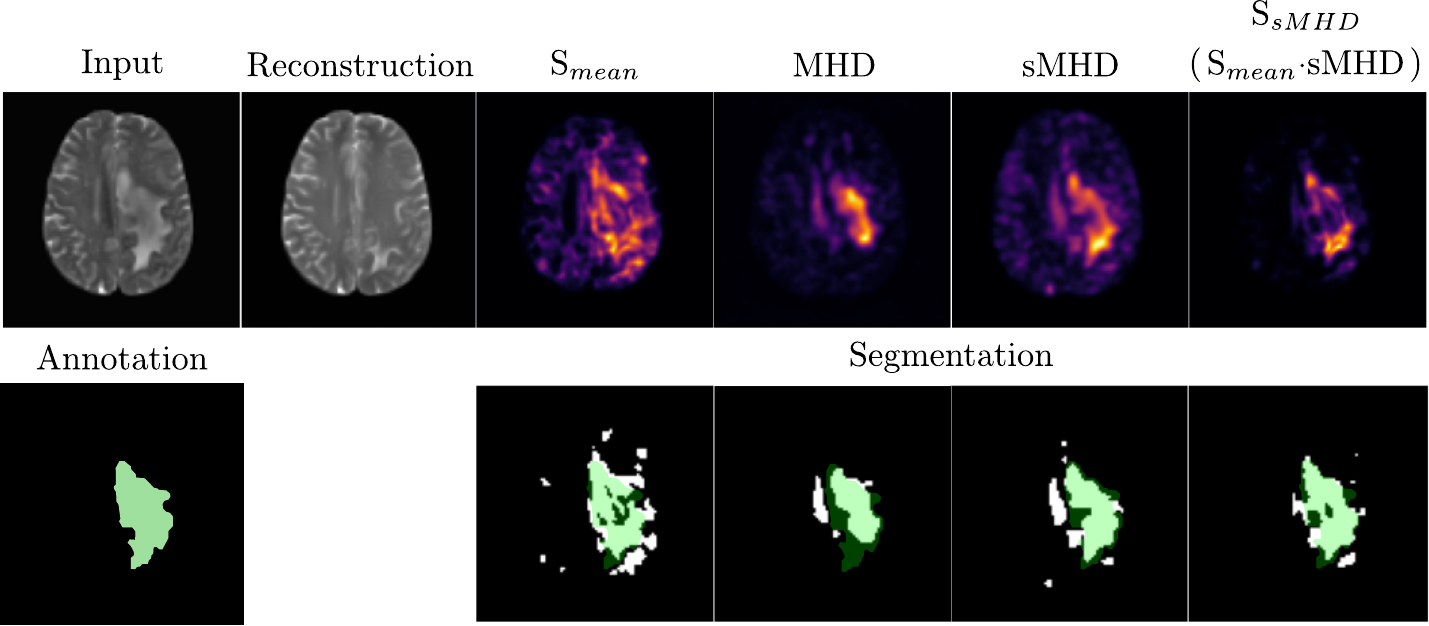} &
    \includegraphics[width=.27\linewidth]{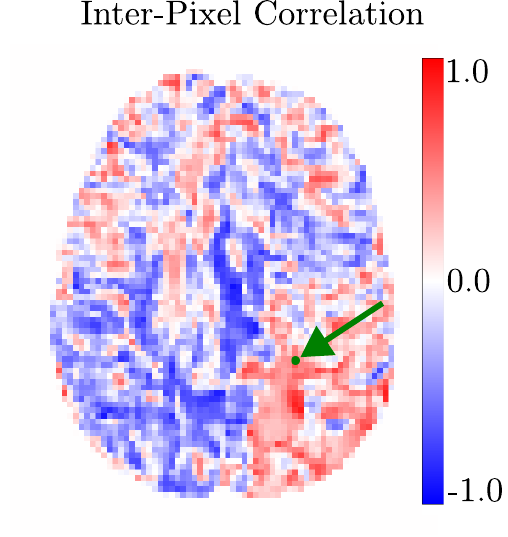}\\
    (a)&(b)\\
    \end{tabular}
    \caption{(a): \textbf{Top row:} input, reconstruction, S$_{mean}$ (SSIM), MHD, sMHD and the final anomaly map are shown for an exemplary image taken from the BRATS data set.
    \textbf{Bottom row:} the ground truth (green) and binarized segmentation maps (white) are shown. Note that the threshold for the segmentation maps is derived by optimizing the Dice score, based on the ground truth. 
    (b): The correlation of one pixel (green arrow) with all other pixels, derived from $\Sigma_{full}$ is visualized as a heatmap.}
    \label{fig:ex_recos}
\end{figure}

We compare the segmentation performance of different variants of our approach to established state-of-the-art baselines. We average the metrics across the five folds and report the mean $\pm$ standard deviation. We initially tested different values for the number of reconstructions N in the range $N=[5,10,...,30]$ and observed a moderate improvement in performance up to $N=10$, after which performance plateaued. Therefore, to balance performance and inference time, we selected $N=10$ reconstructions for each input image. \\\\
The results are shown in Table \ref{tab:segmentation} and  Fig. \ref{fig:ex_recos}. Comparing the baseline models, DAEs exhibit strong segmentation performance for the BRATS data set but are surpassed by cDDPMs for other pathologies in terms of Dice scores. Similarly, feature-based approaches (FAE and RD) perform well on individual data sets but struggle with generalization across all pathologies. Self-supervised approaches (PII and DRAEM) demonstrate poor performance across most data sets. Additionally, the CM method is consistently outperformed across all data sets. Overall, cDDPMs perform robustly across the evaluated data sets while enabling probabilistic sampling of multiple reconstructions. Hence, we consider cDDPMs to generate the pseudo-healthy distributions required for the MHD calculation. Our preliminary experiments indicate that other DDPM variants, such as the baseline DDPMs and pDDPMs, can also be utilized.\\
We find that averaging multiple reconstructions in cDDPMs (S$_{mean}$) enhances segmentation performance across most data sets compared to using a single reconstruction. In contrast to leveraging the MHD with a diagonal covariance matrix (S$_{MHD}$), utilizing the MHD with a full covariance matrix (S$_{sMHD}$) consistently demonstrates improved or competitive performance across all data sets. Notably, compared to the baseline cDDPMs, sampling multiple reconstructions and calculating the sMHD increases the processing time from 0.4 s to 4.9 s per volume. 
\begin{table}[t]
    \caption{Segmentation performance regarding $\lceil$Dice$\rceil$ and AUPRC. The highest values are shown in \textbf{bold}, where \underline{underlines} denote statistical significance ($p<0.05$). S$_{mean}$ denotes the averaging of multiple reconstructions to derive the anomaly map. S$_{MHD}$ and S$_{sMHD}$ denote the use of the MHD either with a diagonal covariance matrix or with a full covariance matrix, respectively. }
     \resizebox{\textwidth}{!}{
    \begin{tabular}{lcccccccc}
        \toprule
         \textbf{Model} & \multicolumn{2}{c}{\textbf{BRATS}}  & \multicolumn{2}{c}{\textbf{ATLAS} }& \multicolumn{2}{c}{\textbf{MSLUB}}  & \multicolumn{2}{c}{\textbf{WMH} }  \\
         \cmidrule(l){2-3} \cmidrule(l){4-5} \cmidrule(l){6-7} \cmidrule(l){8-9} 
          & $\lceil$\textbf{DICE}$\rceil$ &  \textbf{AUPRC} & $\lceil$\textbf{DICE}$\rceil$ & \textbf{AUPRC}& $\lceil$\textbf{DICE}$\rceil$ & \textbf{AUPRC} & $\lceil$\textbf{DICE}$\rceil$ & \textbf{AUPRC}    \\
\midrule
            CM \cite{Saase.2020} & 20.47 $\pm$ 0.22 & 14.03 $\pm$ 0.29 &    12.52 $\pm$ 0.47 &  9.31 $\pm$ 0.69 &  5.24 $\pm$ 0.27 &  2.59 $\pm$ 0.17  & 5.59 $\pm$ 0.09 &   2.70 $\pm$ 0.08 \\
            AE \cite{Baur.2021b} & 36.69 $\pm$ 0.20 & 33.58 $\pm$ 0.29 &    14.03 $\pm$ 0.27 & 11.68 $\pm$ 0.36 &  6.22 $\pm$ 0.05 &  3.55 $\pm$ 0.05 &  9.44 $\pm$ 0.26 &   5.60 $\pm$ 0.21 \\
           VAE \cite{Baur.2021b}&   36.04 $\pm$ 0.91 & 32.84 $\pm$ 1.07 &    14.48 $\pm$ 0.38 & 12.09 $\pm$ 0.41 &  6.33 $\pm$ 0.14 &  3.67 $\pm$ 0.11 &  9.52 $\pm$ 0.23 &  5.71 $\pm$ 0.23 \\
           FAE \cite{meissen2022FAE}&   44.60 $\pm$ 2.17 & 43.75 $\pm$ 0.46 &    17.76 $\pm$ 0.16 &  13.97 $\pm$ 0.10 &  6.85 $\pm$ 0.65 &   4.02 $\pm$ 0.10 &  8.81 $\pm$ 0.38 &  4.97 $\pm$ 0.22 \\
            RD \cite{Deng.2022}&  32.57 $\pm$ 0.15 & 27.13 $\pm$ 0.16 &    19.69 $\pm$ 0.26 &  15.65 $\pm$ 0.20 &   6.48 $\pm$ 0.20 &  3.66 $\pm$ 0.18 &   7.48 $\pm$ 0.10 &  4.22 $\pm$ 0.09 \\
           DAE \cite{Kascenas.2022b}&   62.93 $\pm$ 0.55 & 64.76 $\pm$ 0.79 &    19.42 $\pm$ 0.87 & 17.73 $\pm$ 0.88 &  8.35 $\pm$ 0.45 &  5.64 $\pm$ 0.37 & 11.14 $\pm$ 0.47 &  7.92 $\pm$ 0.55 \\
         DRAEM \cite{Zavrtanik.2021}&  32.75 $\pm$ 3.63 & 26.38 $\pm$ 4.43 &     12.80 $\pm$ 1.94 &  9.63 $\pm$ 1.77 &  5.78 $\pm$ 2.29 &  2.66 $\pm$ 1.14 &  6.25 $\pm$ 1.89 &  3.23 $\pm$ 1.11 \\
           PII \cite{Tan.2021} &      40.83 $\pm$ 2.18 & 36.49 $\pm$ 2.63 &     9.73 $\pm$ 1.89 &  7.26 $\pm$ 1.59 &  9.46 $\pm$ 0.43 &  5.21 $\pm$ 0.33 &  6.59 $\pm$ 1.87 &  3.49 $\pm$ 1.02 \\
          DDPM \cite{Wyatt.2022} &  49.46 $\pm$ 1.56 & 47.57 $\pm$ 1.89 &    15.09 $\pm$ 0.64 & 11.85 $\pm$ 0.47 &  9.97 $\pm$ 0.64 &  6.03 $\pm$ 0.37 & 13.91 $\pm$ 0.37 &  9.15 $\pm$ 0.44 \\
         pDDPM \cite{Behrendt.2023}&   54.26 $\pm$ 0.54 &  53.39 $\pm$ 0.70 &    18.83 $\pm$ 0.38 & 15.92 $\pm$ 0.44 & 10.37 $\pm$ 0.67 &   6.40 $\pm$ 0.51 & 15.31 $\pm$ 0.29 &  10.70 $\pm$ 0.21 \\
         cDDPM \cite{Behrendt.2023c}&   54.39 $\pm$ 0.70 & 54.31 $\pm$ 0.83 &     19.85 $\pm$ 0.90 & 16.99 $\pm$ 0.74 & 11.58 $\pm$ 0.35 &   7.76 $\pm$ 0.30 & 16.03 $\pm$ 0.88 & 12.15 $\pm$ 0.91 \\
         \midrule
   cDDPM S$_{mean}$ &   58.53 $\pm$ 0.48 & 59.14 $\pm$ 0.57 &    21.06 $\pm$ 1.09 & 18.17 $\pm$ 0.93 & 11.75 $\pm$ 0.44 &  7.75 $\pm$ 0.49 & \textbf{17.09 $\pm$ 1.24} & 13.15 $\pm$ 1.25 \\
cDDPM S$_{MHD}$ &    58.47 $\pm$ 0.59 & 61.28 $\pm$ 0.63 &    20.34 $\pm$ 1.26 & 17.51 $\pm$ 1.23 & 12.25 $\pm$ 0.62 &  7.99 $\pm$ 0.69 & 16.82 $\pm$ 1.68 &  13.34 $\pm$ 1.90 \\
  cDDPM S$_{sMHD}$ &  \underline{\textbf{64.72 $\pm$ 0.52}} &\underline{\textbf{ 68.55 $\pm$ 0.63}} &    \underline{\textbf{26.67 $\pm$ 1.61}} & \underline{\textbf{24.61 $\pm$ 1.57}} & \textbf{15.44 $\pm$ 0.85} &\underline{\textbf{ 11.47 $\pm$ 0.79}} & 16.65 $\pm$ 1.45 & \textbf{13.77 $\pm$ 1.57} \\
\bottomrule
    \end{tabular} 
    }
    \label{tab:segmentation}
\end{table}
As illustrated in Fig. \ref{fig:ex_recos} (a), refining the anomaly map of cDDPMs by the sMHD leads to focused anomaly maps. Considering Fig. \ref{fig:ex_recos} (b), we observe non-zero correlations across the entire brain.  Specifically, there exists a symmetric pattern regarding the tumor region with negative correlations in the left hemisphere and positive correlations in the right hemisphere. Exemplary anomaly maps for different models are provided in the supplementary material.
\section{Discussion and Conclusion}
A notable challenge of reconstruction-based UAD methods is their high sensitivity to imperfect reconstructions, often resulting in false positives that impede segmentation accuracy. To address this challenge, we propose to refine anomaly scoring by employing the MHD in the pixel space and identifying anomalies as outliers from pseudo-healthy distributions generated by cDDPMs.\\
Our results (as shown in Table \ref{tab:segmentation}) show that averaging multiple reconstructions from pseudo-healthy distributions (S$_{mean}$) can already improve segmentation performance. This improvement could be attributed to the variability of the noise structure added before reconstruction during the forward process of the cDDPM, resulting in regions with varying levels of complementary information available for denoising.
Notably, applying the MHD with a diagonal covariance matrix (S$_{MHD}$) results in performance comparable to that of averaged reconstructions (S$_{mean}$). In contrast, using the spatial MHD (S$_{sMHD}$) substantially improves the segmentation performance. 
Fig. \ref{fig:ex_recos} (a) illustrates the differences between MHD and sMHD. It can be observed that the MHD is less sensitive to the edges of pathologies compared to sMHD. This indicates that the reconstructions exhibit higher variance in these regions, leading to a smaller weight in the anomaly map. In contrast, the anomaly map derived by sMHD shows improved pathology coverage. Fig. \ref{fig:ex_recos} (b) indicates the presence of inter-pixel correlations across the entire image, ranging from local neighborhoods to global dependencies exhibiting symmetry. However, the MHD with a diagonal covariance matrix does not capture these correlations.
Consequently, the improved performance of sMHD compared to the MHD highlights the importance of considering these dependencies to identify abnormal pixels as outliers. Furthermore, our results indicate that considering the training data as a reference distribution for the MHD, as done in the case of CM \cite{Saase.2020}, is ineffective for segmentation. This finding underscores the importance of constructing a pseudo-healthy reference distribution tailored to each individual test case, which is a key aspect of our approach.\\
In summary, leveraging the sMHD based on generated pseudo-healthy distributions for refining anomaly scoring can enhance the segmentation performance of DDPMs in the context of UAD in brain MRI. While we demonstrate this improved performance for cDDPMs, the baseline DDPMs and pDDPMs can also benefit from the sMHD, underscoring our method's versatility and potential impact in enhancing anomaly detection performance.
A general limitation of UAD is its restriction to binary segmentation and the overall low performance for subtler anomalies, such as those found in WMH or MSLUB data sets. While our approach increases overall performance, it is important to acknowledge the increased computational demand due to the requirement of multiple reconstructions and matrix inversion. Future work could explore efficient approximations or decompositions to enhance the computational efficiency of MHD calculations.
\begin{credits}
\subsubsection{\ackname}
This work was partially funded by grant number KK5208102HV3 and ZF4026303TS9 (Zentrales Innovationsprogramm Mittelstand) and  by the Free and Hanseatic City of Hamburg (Interdisciplinary Graduate School).
\subsubsection{\discintname}
The authors have no competing interests to declare that are relevant to the content of this article.
\end{credits}
\bibliographystyle{splncs04}
\bibliography{Paper-1502}
\end{document}



\setcounter{page}{1}

\begin{figure*}[h]
\centering
        \includegraphics[width=\columnwidth]{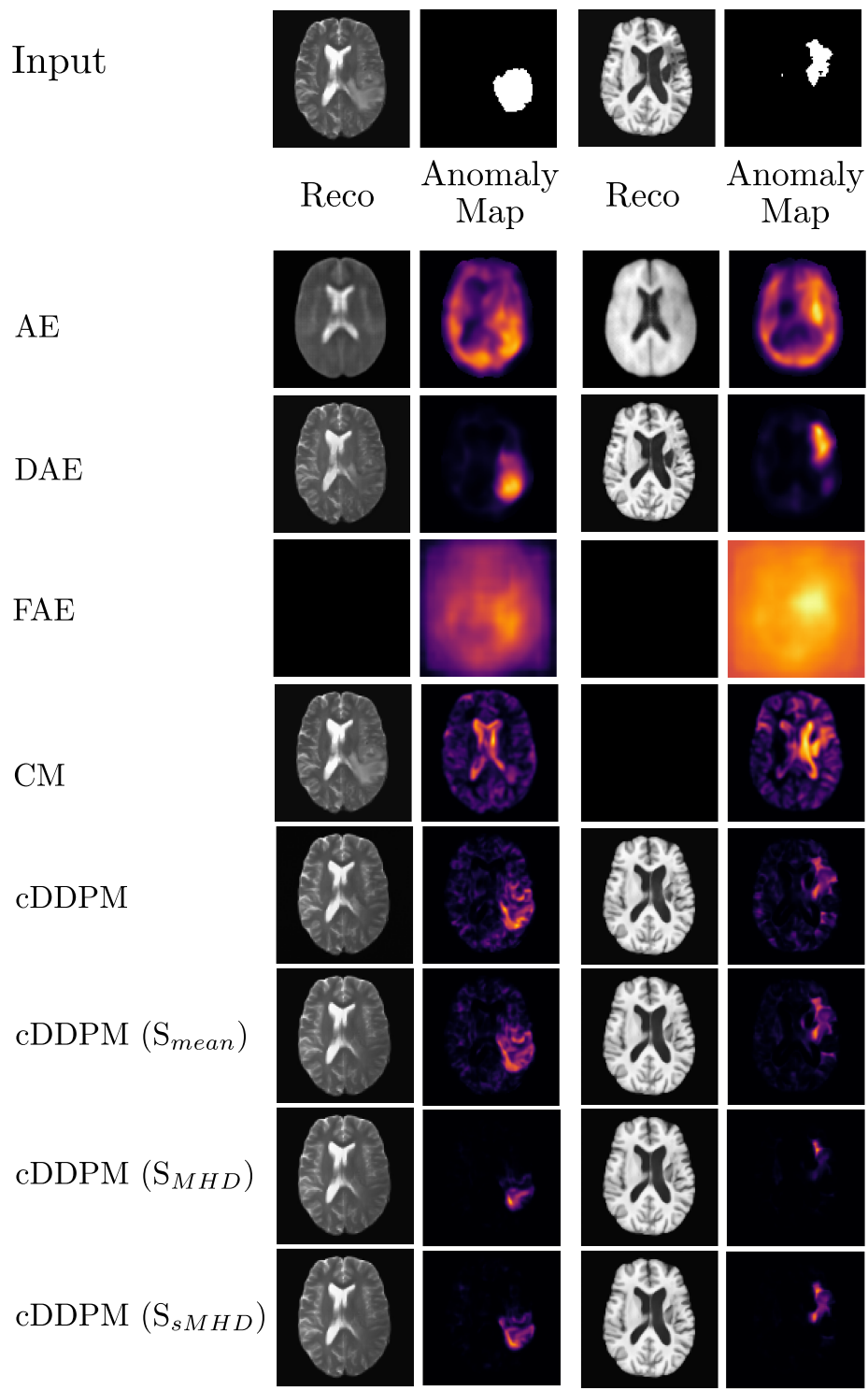}

    \caption{Qualitative comparison of baseline models for pathologies from the BRATS (left) and ATLAS (right) data sets. }
\end{figure*}